# Versatile volumetric additive manufacturing with 3D ray tracing

DANIEL WEBBER,[1,*] YUJIE ZHANG[1], MICHEL PICARD[1], JONATHAN BOISVERT[1], CHANTAL PAQUET,[1] AND ANTONY ORTH[1]

[1]*National Research Council of Canada, Ottawa, Ontario, Canada K1A 0R6*

*\*daniel.webber@nrc-cnrc.gc.ca*

**Abstract:** Tomographic volumetric additive manufacturing (VAM) is an optical 3D printing technique where an object is formed by photopolymerizing resin via tomographic projections. Currently, these projections are calculated using the Radon transform from computed tomography but it ignores two fundamental properties of real optical projection systems: finite etendue and non-telecentricity. In this work, we introduce 3D ray tracing as a new method of computing projections in tomographic VAM and demonstrate high fidelity printing in non-telecentric and higher etendue systems, leading to a 3X increase in vertical build volume than the standard Radon method. The method introduced here expands the possible tomographic VAM printing configurations, enabling faster, cheaper, and higher fidelity printing.

## 1.    Introduction

Volumetric additive manufacturing (VAM) is a new branch of additive manufacturing (AM) in which parts are formed volumetrically instead of the traditional layer-by-layer approach, enabling the manufacturing of new technologies such as micro-optics[1], micro-fluidics [1,2], silicone-based medical devices [3], and rapid production of complex geometries [1-5] Since its inception, tomographic VAM [3] has become the most widely used type of VAM and is based on the core concepts from computed tomography (CT). Here, tomographic projections of the desired object are transmitted by an optical source through a rotating, cylindrical volume of photopolymerizable resin until the delivered light dose meets the gelation threshold of the resin, culminating in the printed part. Due to the volumetric nature of the approach, parts can be quickly formed with smooth surfaces [1-7] without the need for support structures, unlocking printing of new geometries not previously possible.

   The light dose required to print an object is traditionally computed using the traditional Filtered Back Projection (FBP) algorithm [8] used in CT reconstruction. A series of projections of the target object as viewed at different angles is computed. The intensity of each pixel in the projection image represents the line integral along the same angle of the projection viewpoint. This is known as the Radon transform of the target object. Through the central slice theorem, the target object can be reconstructed given enough projections spanning a sufficient range of viewing angles. However, a requisite of this theorem is that the line integrals must be computed over parallel paths. As a result, tomographic 3D printers are constructed to have chief rays parallel to the optical axis (i.e. telecentric) and have a minimal spread of light, called etendue. Several different tomographic VAM printer configurations exist that meet these requirements, and all are based upon illuminating a digital micromirror device (DMD) with an ultraviolet (UV) optical source. One such possible configuration uses a UV laser with telecentric focusing optics and the printing vial immersed in an index matching fluid to simultaneously achieve low etendue and parallel rays in the print volume [1-3]. Another more commonly employed configuration uses a UV LED for illumination, where telecentricity and low-etendue are achieved by either using specialized optics [4-6, 8-11] or by approximation, using standard projection lenses with long projector focus distances [12,13].

The simplest form of a tomographic VAM printer uses the latter configuration and is shown in Fig. 1. It is composed of an off-the-shelf UV projector focusing onto a cylindrical vial mounted to a rotation stage. This configuration has two key advantages over traditional systems: 1) It contains a minimum amount of hardware reducing the complexity and cost. 2) An index-matching bath is not required around the vial further simplifying the system as well as improving its versatility since the immersing bath must have the same index of refraction as that of the photosensitive resin. This avoids the time-consuming process of matching the properties of the bath when printing with new materials. High-fidelity printing in a projector-based VAM without index-matching bath was demonstrated by Orth et al. [12]. Here, they accounted for projector non-telecentricity and refraction at the air-vial interface via a remapping of the Radon coordinates during calculation of the tomographic projections. However, because the method in [12] is only applied within the vial-plane, non-telecentric error along the vial axis is not accounted for, placing a limit on the achievable vertical size of the build volume as well as the minimum tomographic printer system length. This is a consequence of geometry; to maintain a small chief-ray angle with respect to the optical axis, the projector focus distance must be made much larger than the projected image size.

The speed of tomographic printing is partly determined by the amount of light that can be delivered to the absorbing print volume. In a LED-driven projector-based tomographic printer, the large etendue of the LED source requires a small system aperture to achieve a small etendue, making the system optically inefficient. The amount of light can increased via two ways: 1) Increasing the LED current or 2) Increasing the size of the system aperture. The first is normally employed as it maintains the tomographic printing requirements, however light output is typically linear with input current and so the range is limited. Furthermore, due to the optical inefficiency much of the input power is wasted as thermal loss. The more attractive option is to increase the system aperture because the light output scales as the square of the aperture diameter. However, this method has the drawback that the etendue requirement is violated since the projector depth of field is shorted, limiting the resolution and build volume of the printer system [3].

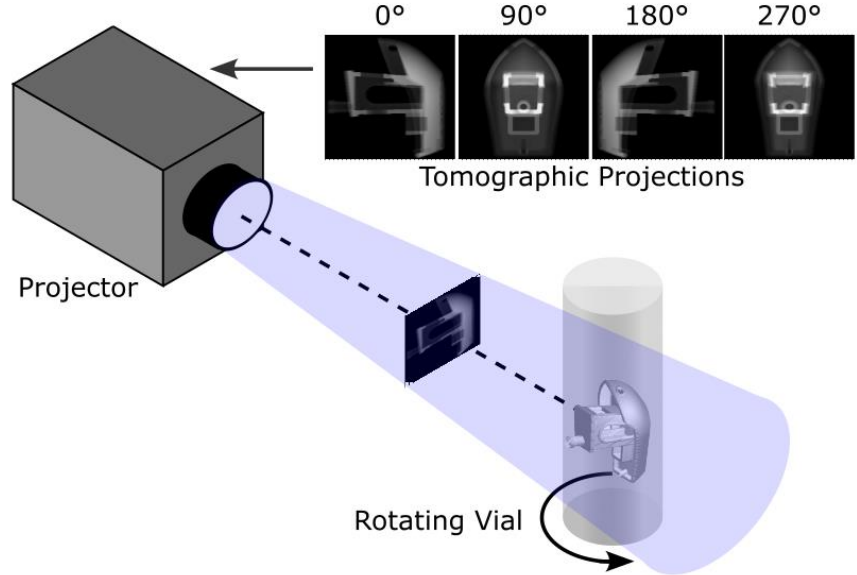

Fig. 1. Experimental setup for a simple tomographic VAM system composed of a UV projector and rotating cylindrical vial. A series of tomographic images are transmitted by the projector through the vial as it rotates. In the diagram tomographic projections for a 3DBenchy print are shown.

Presently, there does not exist a software-only method to account for projector non-telecentricity and etendue in all three dimensions. Computational approaches are attractive because they offer a no-material cost improvement to print fidelity, and reduce the amount of hardware (e.g. cameras, optics) needed which is especially beneficial for applications where size and weight are a premium. In this work, we introduce a new method of computing tomographic projections in VAM. In this method, the tomographic projections and dose simulation are computed using optical rays that are computationally cast in three-dimensions in the printing system, which we refer to as 3D ray tracing (3DRT). This method distinguishes itself from previous Radon-based approaches in two ways: 1) It does not require parallel rays since light dose can be delivered between adjacent layers in the print volume, enabling accurate modeling of non-telecentricity and chief-ray focusing. 2) Our method also does not require that rays are non-diverging within the print volume since etendue can be simulated with the use of additional non-chief rays. To the best of our knowledge, this is the first reported instance of using ray-tracing to directly compute the tomographic projections and dose. Moran et al [14] used a commercial ray-tracing software [15] to examine the optical effects of cylindrical lensing imparted by a vial without index-bath immersion, a vial in an index-matching immersion, as well as the design of a negative lens to compensate for the cylindrical vial lensing. However, they do not create a software-correction for the first two scenarios and instead focus on the design of a physical lens to compensate for the cylindrical focusing.

In Figs. 2(a,c) we show the ray paths for a tomographic printer based on an ideal lens projection system in air (index $n_1$) focused onto a cylindrical vial (index $n_2$) containing photopolymerizable resin (index $n_3$). The chief and marginal rays are plotted as solid and dashed lines respectively. Due to the cylindrical geometry of the vial, strong refraction occurs in the XY plane resulting in chief rays being focused towards the optical axis. Furthermore, due to rotational asymmetry of the cylindrical vial, the system is astigmatic resulting in different foci for the sagittal and tangential planes. This is visually depicted as different points of focus in the XY and YZ planes of Figs. 2(a,c) respectively.

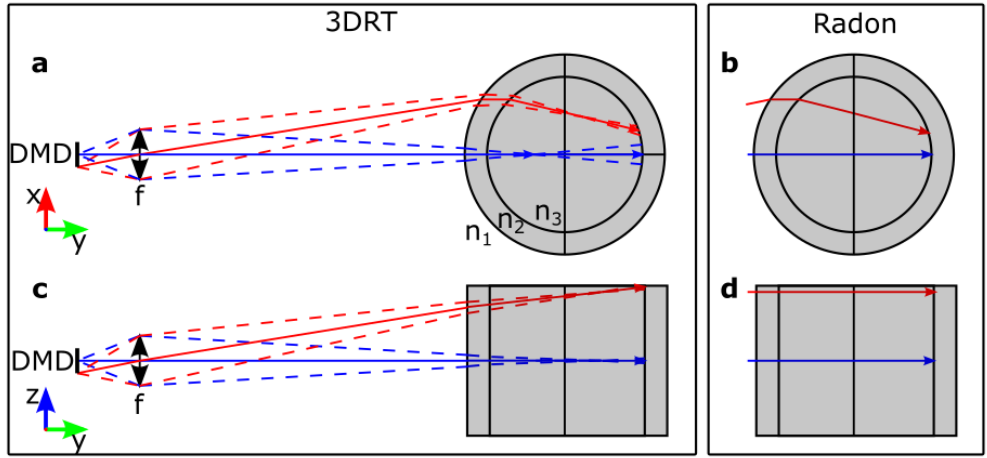

Fig. 2: (a,c) Ray paths for an ideal lens projection system showing the chief (marginal) rays in solid (dashed) passing through a vial of photoresin.. (b,d) Ray paths using the Radon method from [10].

In contrast, rays computed using the Radon approach (Figs. 2(b,d)) from [10] have two important differences. First, only chief rays are considered which is valid only when a low-etendue source is used. Secondly, due to the dimensionality of the approach, rays are approximated as telecentric in the ZY plane (shown as flat rays in Fig. 2d) placing a limitation on the correction for larger chief-ray angles with respect to the optical axis.

In this work, we demonstrate a software based approach to correct for 3D non-telcentricity and larger etendue than what is traditionally used in tomographic VAM. Accurate modeling of the optical rays is used as an *a priori* feedback mechanism to improve print fidelity in a non-idealized tomographic printer. First, we outline the 3DRT method and how it is applied to the two tomographic printing systems used in this work. Next, we demonstrate that 3DRT can compensate for 3D non-telecentricity inherent in a projector-based tomographic printer through comparison of printed parts made with 3DRT and the state of the art software method. We find parts printed without 3DRT degrade rapidly in print fidelity along the vial axis, whereas parts printed with 3DRT obtain good conformity across the entire print volume. This demonstrates that 3DRT effectively increases the vertical build volume of these printers. In addition, we show that 3DRT can be used to print parts in tomographic printers with larger etendue than traditionally used. This observation is especially important, as 3DRT can relax the etendue requirements of tomographic printing and enable faster, more-light efficient systems.

## 2. Materials and Methods

### 2.1 Three-dimensional ray tracing (3DRT)

In this section we describe the method used to determine light ray paths through a tomographic printing system, which we call three-dimensional ray tracing (3DRT). In this work, we considered two types of projection optics attached to a projector containing a UV LED source that is uniformly illuminating a digital micromirror device (DMD) which we refer to together as a pixel array.

The method of ray-tracing used in this work is widely used in engineering and computer graphics applications and so will only be summarized in this report. Ray propagation and refraction at each material interface is computed using the vectorized form of Snell's law [16]. For each pixel in the pixel array, a set of rays are propagated to the first optical element in the system. The direction of each ray is stored as a Cartesian unit vector, and the direction of the chief ray is determined by the location of the aperture stop in the optical system. The direction of the non-chief rays is determined by the size of the aperture stop. Here, we use hexagonal filling to define non-chief ray locations on the aperture stop. We then compute the direction of each non-chief ray based on these locations.

At the intersection of the light-ray and each optical element surface, the direction of the refracted ray is computed and the ray is propagated to the next surface in the system. This process is repeated until the rays have intersected the final surface of the system, which in this work is the inner-diameter of the vial furthest from the projector. The ray coordinates within the vial are used to compute a set of Cartesian indices of intersection between the discrete voxel array and each ray passing through the print volume. These indices of intersection are used in computing both the tomographic projections as well as the light dose delivered which is discussed later.

## 2.2 STL Conversion to Voxel Array

The STL file representing our desired part is first converted into a logical 3D voxel array using the Mesh Voxelisation package [17] in MATLAB, where 1 or 0 represent the presence or absence of the part. Next, the data is converted into single-precision floating-point data type. In Fig. 3a is shown a voxelized 3DBenchy object. Calculation of the tomographic projections and dose is computationally intensive. To increase the calculation speed for the two printers studied in this work, the target object was down-sampled by 4 X 4 X 4 projector pixels.

## 2.2 Anti-aliasing

A super-sampling method was implemented to avoid aliasing artefacts so that voxels at the ray location, as well as adjacent voxels are included in the tomographic projections and dose calculations. First, the voxel pattern in Fig. 3a is upsampled by a factor of 2 so that each voxel now extends to a 2 X 2 X 2 region, as shown in In Fig. 3b. For a ray encountering this region and located at (x0,y0,z0), the ray will interact with the nearest integer voxel, as well as the next-nearest integer voxels. The strength of the interaction, or weight, of the ray to each voxel is determined by the distance between the ray and each voxel $(x_v, y_v, z_v)$, i.e.

$$[d_x, d_y, d_z] = |[x_0, y_0, z_0] - [x_v, y_v, z_v]|$$

$$weight = |(1 - d_x) * (1 - d_y) * (1 - d_z)|$$

## 2.3 Tomographic Projection Calculation using Radon Method

Tomographic images using the Radon method were computed in MATLAB following the procedure outlined in [12] and is briefly summarized here. The built-in Radon function in MATLAB [was used to compute the Radon transform for each z-slice in the voxelized object. The resulting sinograms were then passed through a Ram-Lak filter in the Fourier domain (with negative intensity values clipped to 0) followed by a coordinate system remapping to account for refraction at the air-vial interface as well as non-telecentricity of the optical source.

## 2.4 Tomographic Projection Calculation using 3DRT

Tomographic images using 3DRT were calculated using the same methodology as the Radon-based approach but for line-integrals in three-dimensions and for multiple rays per pixel. In Fig. 3c is a diagram showing the propagation of a single ray through a small section of the target dose in Fig. 3a. As the ray propagates through the voxel array containing the target dose (depicted by the blue-outlined voxels), it intersects all black-outlined voxels due to the anti-aliasing method described above. The intensity of the pixel from which this ray originated from is given by the sum of all voxels that the ray intersects with the target dose, shown as the solid blue voxels. This is repeated for N rays from each pixel (typically N = 1 is used for a low etendue source), for each pixel in the pixel array, and subsequently for 360 angular samples of the target dose.

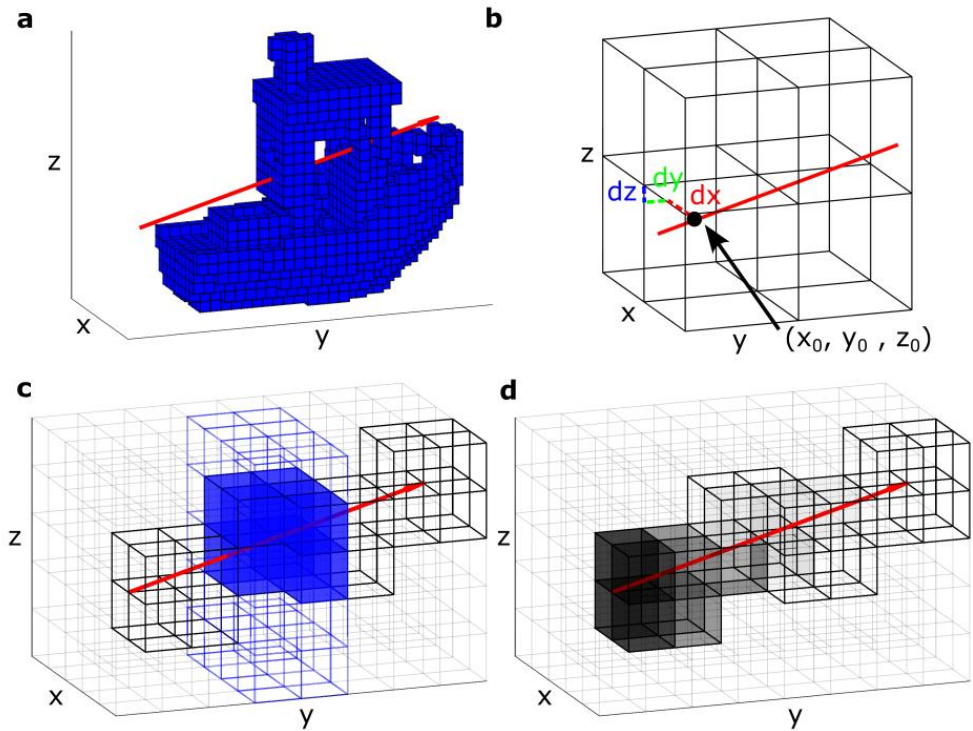

Fig.3: a) Image of a single ray (red) passing through a voxelized 3D Benchy. b) Diagram showing the method of anti-aliasing used in this work. c) The tomographic projections were computed for each ray (red) by determining the intersection between the target dose voxels(blue-outline) and the voxels the ray encounters (black-outline). The intersection voxels are shown as solid blue. d) The tomographic dose for each ray was computed by propagating the ray through all voxels the ray encounters. Here, the darker the shading the higher the received light dose. Due to optical absorption, voxels nearest the source (e.g. projector) receive a higher light dose.

*2.5 Dose Simulation*

The rays used to calculate the tomographic projections are also used to simulate the delivered dose and is described as follows. Tomographic projections covering 360 degrees in 1 degree increments were transmitted through the simulated print volume. For each tomographic image, a ray from each pixel was cast through the print volume along the pre-determined ray path, as shown in Fig. 3d. For each voxel the ray intersects, a dose is added to the voxel that is the product of the tomographic pixel intensity (determined from the tomographic projection calculation shown earlier), the scaling weight due to the anti-aliasing method, the transmission loss at the print volume interface, and the transmission loss due to optical absorption through the print volume. Shown in Fig 3d is shown the corresponding dose calculation for the tomographic projection calculation shown in Fig 3c. The magnitude of the delivered light dose is shown by the voxel shading, where darker corresponds to more dose. Due to optical absorption, voxels nearest the optical source receive more dose. Optical transmission at the print-volume interface is calculated using the Fresnel coefficients for unpolarized light, and attenuation within the volume is computed using Beer-Lambert absorption. In the case of simulating a system with finite etendue, the above approach is repeated for the number of rays cast from each pixel. Finally, the above is repeated for all tomographic projections culminating in the final dose delivered to the print volume.

### 2.6.1 Non-telecentric tomographic printer

The first tomographic printer used in this work is an off-the-shelf Digital Light Innovations CEL5500 projector (DMD pitch 10.8 $\mu m$). It has a throw ratio of 1.8, with a projection focus distance of 99.5 mm corresponding to a pixel size of 0.054 mm. As the composition of the exact projector optics is unknown, we instead modeled it as an ideal lens (focal length 16.58 mm) coincidentally positioned with the aperture stop of the system. A vial (outer diameter (OD) = 25 mm, inner diameter (ID) = 23.6 mm) is mounted to a rotation stage (Physik Instrumente M-060) located at the origin. Based on the downsampling above, the minimum voxel size is 0.216 mm. In Fig. 4 is shown the chief-ray paths for the idealized system.

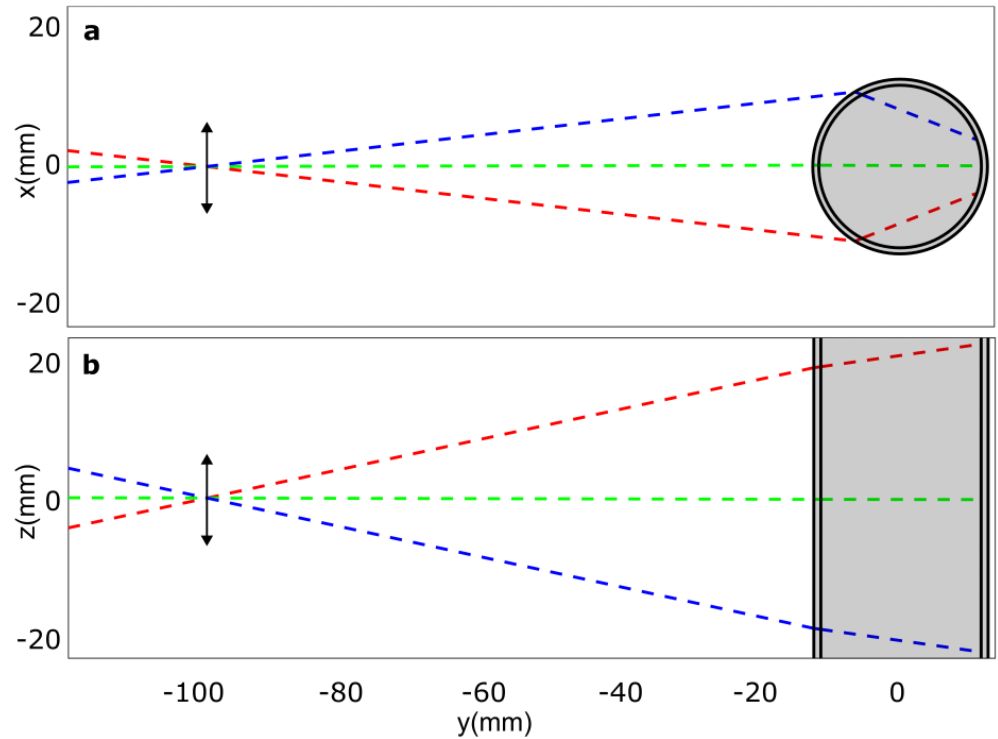

Fig. 4: Ray paths in the non-telecentric VAM printer for a) elevation and b) plan views respectively. Rays emerge from the pixel array at the left-hand side of the image, pass through the idealized projection lens (depicted as a double-sided arrow) and transmit to the vial containing photopolymerizable resin.

### 2.6.2 Telecentric tomographic printer

The second tomographic printer used in this work also uses an off-the-shelf digital light innovations CEL5500 projector (DMD pitch 10.8 $\mu m$) but with the projection lens replaced with a telecentric lens system. The printer is composed of a DMD chip (located at y = -294 mm) which serves as the pixel source, and two plano-spherical lenses (Thorlabs LA1608) with an iris (2.5 mm or 16 mm) placed midway (y = -150 mm). Due to the optical arrangement, the system is both object and image-space telecentric with unity magnification resulting in a pixel size equal to the DMD pitch (10.8 $\mu m$). A vial (OD = 8 mm, ID = 6.6 mm) is mounted to a rotation stage (Physik Instrumente M-060) located at the origin. In Fig. 5 is shown the ray paths for the printer with the 16 mm aperture. Based on the downsampling above, the minimum voxel size is 0.0432 mm.

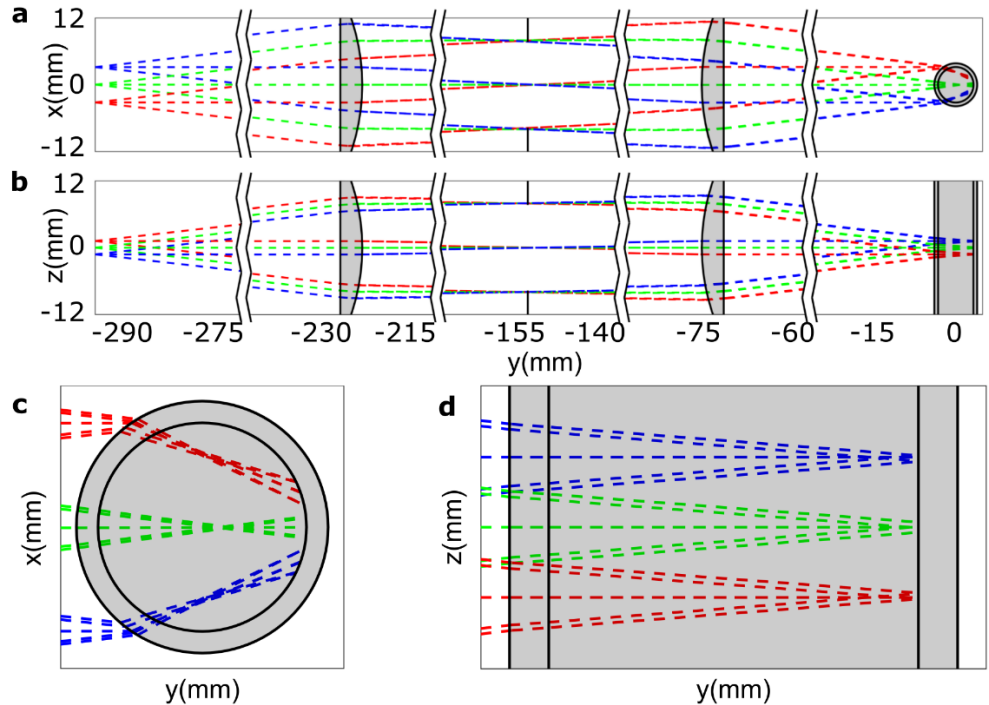

Fig. 5: Ray paths in the (a) XY and (b) YZ planes for a tomographic printer with a telecentric projection system and a 16 mm aperture placed midway between the two lenses. The different colors represent ray paths from different pixels in the pixel array. Graph breaks (depicted as white wavy lines) are placed between optical components for clarity. (c) Magnified view (5X) of rays within the vial in the XY plane. (d) The same as (c) except in the YZ plane and 8.5X magnification.

### 2.7 Optimization

In this work, we use the object-space model optimization (OSMO) feedback algorithm from Rackson et al. [10] in conjunction with 3DRT to produce optimized tomographic projections. We replaced the Radon and the inverse Radon functions in [10] with custom functions that compute the tomographic projections and dose computation as described in Sec. 5.1. Furthermore, to improve the contrast of our tomographic projections relative to the projector black level, we have added a histogram equalization step after the final projection calculation but before the final dose simulation. Projection values exceeding three standard deviations from the mean projection value are clipped and then rescaled within the 8-bit greyscale range of the image being exported to the projector. Optimization was terminated when the absolute relative error between the target dose and the simulated dose was below 0.5 percent.

### 2.8 Vial Alignment

Both systems had resin-filled vials centered on a rotation stage with custom designed vial holders. The position of the vial within the field of view of the projector was aligned as follows; The transverse location of the vial was set by first projecting a vertical line from the projector and marking its location on a piece of paper past the rotation stage. Next, a vial with photopolymerizable resin was placed into the beam path, and translated transversely through the beam until the transmitted vertical line matched the original marked location. The position of the vial along the optical axis of the projector was set by visually inspecting the projected image, and aligning the center of the rotation stage to the point of best image focus.

### 2.9 Imaging

Both tomographic printers used in this work used the optical scattering tomography (OST) modality [13] to inspect printing behavior as well as determine print completion. In both systems, a collimated red 625 nm LED source (Thorlabs #M625L4) was mounted vertically and directed towards the center of the vial. The aperture on the light source was reduced to avoid light scatter on the interior edges of the vial. The OST imaging systems were different to accommodate the different printing resolutions of both systems and are described below.

In the non-telecentric printer, the imaging camera is a FLIR USB3 Grasshopper (GS3-U3-23S6M) with a c-mount projection lens (Edmund Optics 25 mm/F1.8 #86572). The camera was set to use 4 X 4 pixel binning with an imaging resolution of 0.155 mm. The imaging system in the telecentric printer is also telecentric and is composed of two plano-convex lenses, focal lengths 100 mm and 125 mm, with the latter closest to the vial. The imaging camera is a FLIR USB3 Grasshopper (GS3-U3-23S6M). The imaging resolution is 7.4 $\mu$m. In both systems, the camera gain and exposure time was adjusted provide good OST contrast without saturation.

### 2.10 Post-Processing

Once the parts were printed, they were removed from the vial and placed in a glass dish filled with isopropyl alchohol (IPA) to soak for 15 minutes. Then, they were removed and placed in a vacuum chamber and pumped for 5 minutes. Finally, while under vacuum, the parts were exposed to a 405 nm flood illumination for 5 minutes.

### 2.11 Resin

The resins used in this work were prepared mixing diurethane dimethacrylate (DUDMA) with poly(ethylene glycol) diacrylate (Mn=700 g/mol, PEGDA 700) in a 8:2 wt ratio. Different photoinitiators were used for the non-telecentric and telecentric systems. For the non-telecentric printer, the resin was mixed with camphorquinone (CQ) and ethyl 4dimethylaminobenzoate (EDAB) at equal concentrations of 7.8 mM. For the telecentric printer, the resin was mixed with ethyl (2,4,5-trimethylbenzoyl) phenylphosphinate (TPO-L) at a concentration of 3.38 mM.

PEGDA 700, CQ, and EDAB were purchased from Sigma Aldrich, DUDMA from EssTech Inc., and TPO-L from Oakwood Chemical. After mixing, the resin was transferred to open top vials of outer diameter values of 25 mm and 8 mm for the non-telecentric and telecentric printers respectively. The vials were kept at room temperature in a dark storage container until all air bubbles in the resin were eliminated (by visual inspection) and to allow the resin to reach room temperature.

The refractive index of the liquid resin was measured to be 1.49969 at 405 nm using a Schmidt-Haensch ATR-BR refractometer. The room temperature viscosity of the DUDMA resin was measured to be 1100 cp using a Brookfield DV-III Ultra Programmable Rheometer.

## 3.    Results

### 3.1 Non-telecentricity correction

We validate our 3D non-telecentric correction using the non-telecentric tomographic printer on a custom part and show the results in Fig. 6. The part was comprised of 9 parallel fins (1 mm thick, 2.25 mm period) oriented normal to the axis of the vial. The part, as well as its location within the vial is shown in Fig. 6a. The object was printed on the axis of the vial with the base (co-located with the origin in Fig. 6a) positioned on the optical axis of the projector. Side and frontal views are also shown in Fig. 6b, 6c respectively.

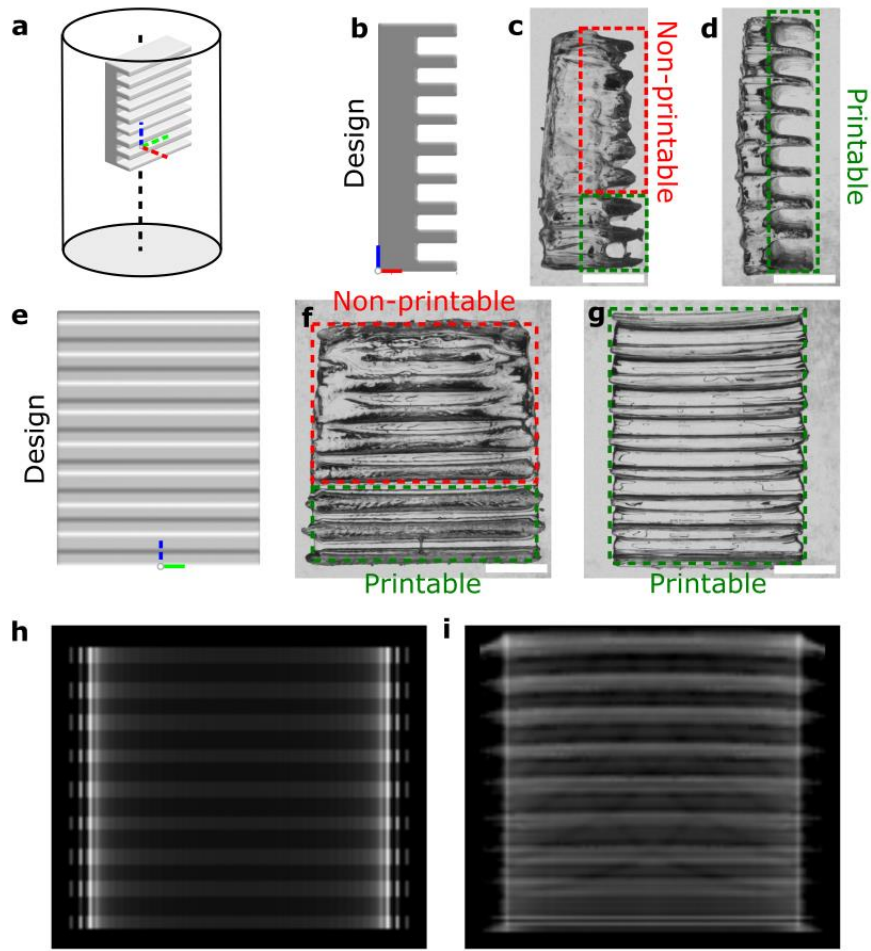

Fig. 6: a) A custom part located at its printing position within the vial. b) Side view of object in (a). c) Image of side of printed object without correction for 3D non-telecentric projection. d) As in (c) but with 3D non-telecentric correction. e) Front-side view of object in (a). f) Image of front-side of printed object without correction for 3D non-telecentric projection. g) As in (c) but with 3D non-telecentric correction. h) One of the tomographic projections calculated using the Radon method used to print the object shown in (c,f). The same as in (h) but with 3D non-telecentric correction and used to print the parts shown in (d,g). The Cartesian coordinate system is denoted by the red, green, and blue colored lines. All scale bars correspond to 5 mm.

The part was printed with 3DRT and using the Radon method. Optical images of these parts are shown in Fig. 6c, 6d and Fig. 6f, 6g respectively for front and side views. For both parts, print termination was determined when all 9 fins formed in the part as viewed using OST. For the parts printed without correction, we see a drop-off in part quality for increasing distance from the projector optical axis, resulting in only 3 of 9 fins printing correctly. We refer to the regions where the fins form and do not form as printable and non-printable respectively in Fig. 6c, f. The front view of the uncorrected part (Fig. 6(f)) most clearly shows this point, as the parallel fins are only resolvable for the portion of the part printed along the vial axis. Regions of the part printed away from the vial axis show a thickening of the fins to the point that they are unresolvable, or non-printable, as can be more easily seen from the side view (Fig. 6(c)). During printing, it was observed that the fins nearest the optical axis formed first. As a result, once the upper-most fins formed the bottom fins became overexposed resulting in a general thickening of the part, as shown by the tapering of the object within the non-printable region of

Fig. 6c. This is a direct consequence of non-telecentricity, as the intensity of light (and dose delivered) per unit voxel will decrease for increasing chief-ray angle with respect to the optical axis. This can be most easily seen by comparing the tomographic projections calculated using the Radon method (Fig. 6h) and with 3DRT (Fig. 6i). Because the Radon method is 2D, the intensity of each slice is constant as a function of vertical position. In contrast, in the projections calculated with 3DRT, we see an increase in average intensity as a function of vertical position. Furthermore, whereas all fins are flat in the Radon projections, the fins in the 3DRT projection are curved further corresponding to the non-telecentric distribution of light rays within the print volume.

In contrast, the part printed with 3DRT shows good conformity to the part, with all 9 fins correctly printed as observed in the printable regions of Fig. 6d, 6g. Further, we see that the solid rectangular base the fins are printed on has a uniform thickness indicating that correction has compensated for the reduction in applied dose caused by chief-ray divergence. A slight bending in the upper-most fins is observed which may be corrected with further optimization of the tomographic projections using the OSMO algorithm.

The increase in vertical build volume using 3DRT over the previous Radon method can be quantified by the number of fins correctly printed using both methods. Since only 3 of 9 fins correctly formed with the Radon method, whereas with 3DRT all fins corrected formed, we estimate that 3DRT results in at least a three-fold increase in print volume to a vertical build size of 38 mm in our printer configuration. Here, the vertical build size was limited by the height of the projector image within the print volume.

### *3.2 Etendue Correction*

The impact of astigmatism and other etendue-related effects on tomographic print quality is demonstrated in Fig. 7. A custom part (model shown in Figs. 7(a,b)) was printed using aperture sizes of 2.5 mm and 16 mm corresponding to a full axial beam divergence of 1.9 and 12.2 degrees respectively. In the 3DRT experiment, 13 rays per pixel were used to account for the effect of etendue.  In Fig. 7 (c) is shown a microscope image of the part printed using the Radon method with a 2.5 mm aperture. The letters in the print are clearly resolvable, and can be more clearly seen in the OST image shown in Fig. 7 (f). Conversely, the part printed using the Radon method with 16 mm aperture (Fig 7 (d,g)) has decreased print fidelity, as evident by the thickening of letters as well as the partial forming of the letter "R". In Fig. 7 (e,h) is a bright-field microscope image and OST snapshot of the part printed using 3DRT and a 16 mm aperture. All letters are clearly formed with good conformity to the target model. In particular, the letter "R" is fully formed and the letters "R" and "C" are clearly separated.

The impact of etendue on the tomographic projections can be more easily observed in Fig. 7 (i,j), where tomographic projections computed using the Radon method and 3DRT are shown. Including etendue in the projection calculation results in a redistribution of intensity as well as an overall blurring.

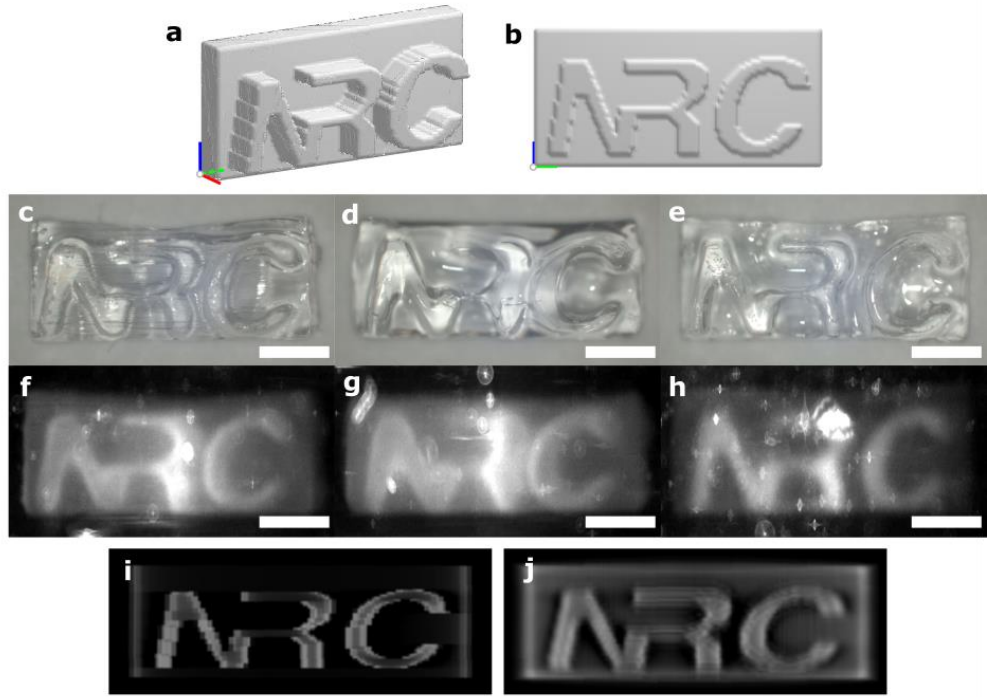

Fig. 7: Improving printing accuracy using 3DRT for different optical aperture size. (a) Digital model of the part. (b) Side view of model. (c) Microscope image of part with 2.5 mm aperture using the Radon method. (d) Same as in (c) but with a 16 mm aperture. (e) Same as in (d) but with correction using 3DRT. (f-g) Video snapshot of the optical scattering tomography signal during printing of the parts in (c-e) respectively. (i,j) Tomographic projections computed with the Radon method and 3DRT respectively. All scale bars are 1 mm.

## 4.    Discussion

We have introduced a new method of computing projections in tomographic VAM By modeling optical rays in 3D, the tomographic projection and delivered dose can be accurately determined enabling new printing configurations not previously possible using conventional Radon-based techniques.

We have demonstrated that 3DRT can increase the build volume of a tomographic printer by maintaining print precision over a larger vertical extent. Analogous to a non-telecentric VAM printer but in the imaging domain, cone-beam CT suffers from undersampling artefacts due to divergence of the cone beam. This causes voxels nearest the source to record more rays than those furthest, resulting in a linear decrease in rays per voxel [18]. This is applicable along the vertical dimension of non-telecentric VAM. Since all regions of our part printed with good fidelity, this indicates that 3DRT accurately accounts for chief-ray divergence along the vial axis. Although we were limited by the vertical size of the tomographic projections at the vial, we anticipate that the build volume could be expanded further to larger chief-ray angles for the same feature size. We investigated this claim by simulating the delivered light dose using 3DRT for a taller version of the custom part in Fig. 6. Shown in Fig. 8a is a digital model of the part as positioned with the vial. The number of fins was increased from 9 to 18 so that the part was twice as tall as the original part. As a result, the maximum chief-ray angle with respect to the optical axis increased from 8.2 degrees to 16.4 degrees. The fin thickness and period was the same as the original part so that the size and shape of features would remain constant. Here, we used the same printer specifications as the non-telecentric printer described in Sec. 2.6.1 except that the number of vertical pixels in the pixel array was increased to account for the larger vertical build size. The simulated dose using 3DRT is shown in Fig. 8c. We find that the dose closely matches the digital model of the part, and that all fins are clearly formed with no observable degradation in quality for increasing vertical height. This is more easily seen by comparing the side view of the custom part and the simulated dose, as shown in Fig. 8b and Fig. 8d respectively. Based on these simulation results, 3DRT could potentially offer a 6X improvement in vertical build size as compared to the Radon approach.

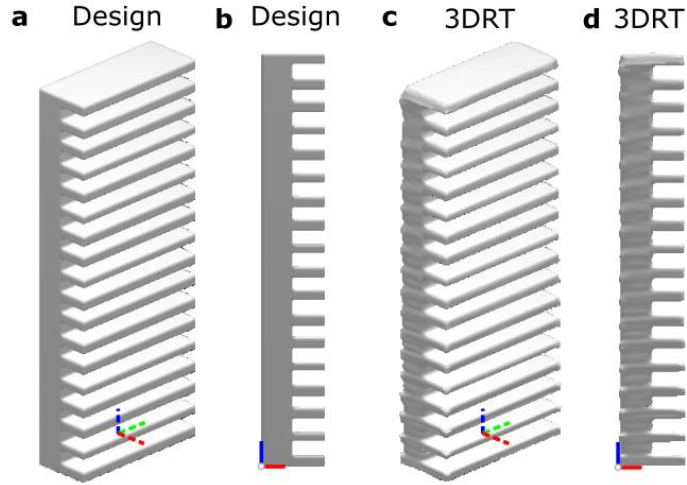

Figure 8: Simulation results of a custom part using a modified version of the non-telecentric tomographic printer. a) Dimetric and b) side view of the design model. The simulated dose computed using 3DRT are shown for the same viewpoints in (c,d) respectively.

As presented in the results, 3DRT enables tomographic printing using optical systems with larger etendue than traditionally used because it directly models etendue-based effects in the computation of the light dose. Within the paraxial limit, for which the printers studied here are applicable, the etendue of a pixel within the print volume is equal to the product of the pixel area and the solid angle of the pixel extent. For the telecentric printer with 16 mm aperture used here (pixel size = 10.8 $\mu m$, full axial beam divergence = 12.2 degrees), assuming a conical light cone ($\Omega = \pi\theta^2$ where $2\theta$ = full axial beam divergence) [19], we obtain an etendue of 4.153E-6 sr mm^2. This value is over an order of magnitude larger than other reports on projector-based tomographic VAM [6,12,13].

An earlier study by Rackson et al. [10] investigated the impact of finite etendue on print quality but in an approximated fashion by modeling beam divergence as a 2D Gaussian [10]. This is approximately valid in the case of parallel rays passing through the write volume that originated from a Gaussian optical source (e.g. a Gaussian-beam laser). However, because their method is only applied in 2D due to the constraints of the Radon transform, the width of the light beam along the vial axis is not accounted for. Further, this approach is not valid for non-parallel optical configurations (e.g. non-telecentric projection or systems without an index-matching bath) and systems where the light cone about each ray is not Gaussian.

We note that existing empirical feedback mechanisms utilizing camera capture hardware [3] could in principle be used to correct for etendue and non-telecentricity. Currently, the rapid rate of resin gelation combined with processing speed limitations requires that these methods to be done sequentially; the object is printed and gelation is recorded. Then, using this information the dose is adjusted so that the part is printed correctly. 3DRT is advantageous because it is a software-based technique and does not require physical feedback (i.e. camera capture) to produce optimized projections.

Although we have shown that 3DRT improves print fidelity in both telecentric and non-telecentric printing systems without the need for an index-matching immersive bath, this method could possibly be used in unconventional printing configurations such as the tomosynthetic geometry in Rackson et al. [10]. 3DRT also enables computation of tomographic projections for systems using multiple writing wavelengths since ray propagation is computed separately for each wavelength. This would be attractive in systems utilizing multiple photoinitiators with different activation wavelengths, such as demonstrated by Wang et al. [9] in controlling material stiffness in printed parts.

We acknowledge that there is a tradeoff in the additional printing performance and that is through the expense of software computation. Currently, computation is performed in MATLAB on a laptop with a four-core i7-4800MQ processor, 24 GB memory, resulting in a non-optimized implementation. Computation of the tomographic projections alone takes between 10 to 100 times longer than a comparable Radon approach, and this differential increases for modeling systems with large etendue necessitating multiple casted rays per pixel. It is anticipated that significant speed gains could be made with GPU implementation. The parallel nature of ray-tracing is widely known to benefit from the use of GPU acceleration (as observed in dedicated ray tracing programs such as FRED by Photon Engineering [20]).

## 5. Conclusion

Projector-based tomographic printers must conform to strict optical specifications to yield high fidelity parts. This is a result of the Radon method to compute the tomographic projections which assumes parallel rays do not diverge when transiting the print volume. 3D image-space non-telecentricity inherent to all projectors as well as larger etendue deviate from these assumptions, leading to undesirable artefacts in the printed parts.

To overcome the limitations of the Radon approach, we have developed a ray-tracing approach to cast rays in 3D so that both tomographic projections as well as the delivered light doses could be computed. Using this software we were able to demonstrate correction of 3D non-telecentricity, resulting in a 3X increase in build volume in a tomographic printer without a change in hardware. Further, using the same software we also demonstrated that high fidelity printing can be achieved in a system with larger etendue than traditionally used. We anticipate the results shown here will enable a broader range of optical configurations for tomographic printers, potentially paving the way to lower-cost printers and commercialization.

**Funding**

Funding provided by the National Research Council Canada.

**Disclosures**

The authors declare the following financial interests/personal relationships which may be considered as potential competing interests: Daniel Webber, Antony Orth, Yujie Zhang, Michel Picard, Chantal Paquet, Jonathan Boisvert, has patent pending to National Research Council Canada.

**Data Availability**

Data available upon reasonable request.

**Acknowledgements**

The authors would like to thank the anonymous reviewers for their time.